\documentclass[%
reprint,
amsmath,amssymb,
aps,
superscriptaddress,
]{revtex4-2}

\usepackage{graphicx}
\usepackage{dcolumn}
\usepackage{bm}
\usepackage{color}

\begin{document}
	

\title{Theory of $\chi^{(2)}$-microresonator-based frequency conversion}

\author{Yun Zhao}
\affiliation{Department of Electrical Engineering, Columbia University, New York, NY 10027, USA}
\author{Jae K. Jang}
\affiliation{Department of Applied Physics and Applied Mathematics, Columbia University, New York, NY 10027, USA}
\author{Yoshitomo Okawachi}
\affiliation{Department of Applied Physics and Applied Mathematics, Columbia University, New York, NY 10027, USA}
\author{Alexander L. Gaeta}
\email{a.gaeta@columbia.edu}
\affiliation{Department of Applied Physics and Applied Mathematics, Columbia University, New York, NY 10027, USA}
	



\begin{abstract}
	Microresonator-based platforms with $\chi^{(2)}$ nonlinearities have the potential to perform frequency conversion at high efficiencies and ultralow powers with small footprints. The standard doctrine for achieving high conversion efficiency in cavity-based devices requires ``perfect matching'', that is, zero phase mismatch while all relevant frequencies are precisely at a cavity resonance, which is difficult to achieve in integrated platforms due to fabrication errors and limited tunabilities. In this Letter, we show that the violation of perfect matching does not necessitate a reduction in conversion efficiency. On the contrary, in many cases, mismatches should be intentionally introduced to improve the efficiency or tunability of conversion. We identify the universal conditions for maximizing the efficiency of cavity-based frequency conversion and show a straightforward approach to fully compensate for parasitic processes such as thermorefractive and photorefractive effects that, typically, can limit the conversion efficiency. We also show the design criteria that make these high-efficiency states stable against nonlinearity-induced instabilities.
\end{abstract}



\maketitle

\begin{figure}
	\centering
	\includegraphics{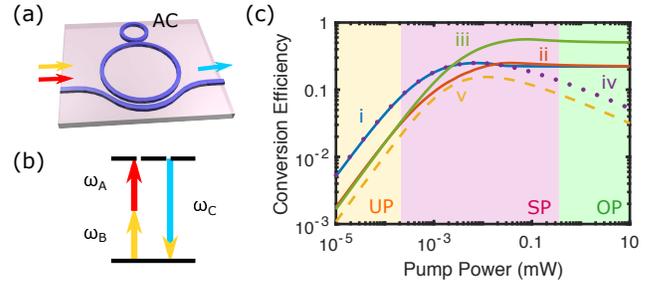}
	\caption{(a) Illustration of a microresonator structure for cavity-based frequency conversion. An auxiliary cavity (AC) can be used for additional tunability. (b) Energy diagram of SFG. The process corresponds to SHG when the pump and signal are degenerate. (c) Numerical simulation of conversion efficiency in a PPLN microresonator with an intrinsic linewidth of 250 MHz. The pump detuning is optimized for each power level except for trace (iv) which corresponds to zero pump detuning. UP, under-pumped regime, SP, saturation regime, OP, over-pumped regime. Traces (i), (ii), (iv) and (v) represent critically coupled cavities and trace (iii) represents overcoupled cavities. In addition, traces (ii), (iii), and (v) represent resonance mismatched cavities. For (i), (ii), (iv) and (v) $\alpha_A = \alpha_C =2\pi\times$500 MHz, and for (iii),  $\alpha_A = \alpha_C =2\pi\times$1 GHz. For (i) and (iv), $\delta_m$ = 0, and for (ii), (iii), and (iv) $\delta_m = 2\pi\times$500 MHz.}
	\label{figScheme}
\end{figure}

Nonlinear frequency conversion is an important tool for realizing broadband coherent sources and for many optical applications such as telecommunications, spectroscopy, and atomic physics. Microresonators [Fig. \ref{figScheme}(a)] have emerged as a powerful platform for frequency conversion due to their small mode volumes, high quality factors ($Q$), and flexible dispersion engineering. Recently, efficient frequency conversion with low pump powers has been demonstrated in several $\chi^{(2)}$ microresonator-based platforms \cite{Furst_PRL_2010,Guo_Optica_2016,Bruch_APL_2018,Logan_OE_2018,Chen_Optica_2019,Lu_Optica_2019,Lu_Optica_2020,Lu_NatPhot_2020,Chen_arXiv_2021}. While these approaches show promise, achieving conversion efficiencies exceeding 50\% remains experimentally challenging. Previous analyses on cavity-based frequency conversion assumed ``perfect matching'' conditions where all frequency components are on-resonance and precisely phase matched \cite{Ilchenko_PRL_2004,Rodriguez_OE_2007,Sturman_JOSAB_2011,Bi_OE_2012,Huang_OL_2013,Strekalov_NewJPhys_2014,Breuning_LaserPhotRev_2016}. It is believed that the violation of such condition results in a reduction of conversion efficiency. However, in practice, the perfect matching condition can be difficult to meet in microresonators, since the resonance frequencies cannot be independently tuned, and their exact locations are sensitive to nanometer scale structure size variations. Furthermore, the cavity resonances can be shifted by other processes, such as thermorefractive and photorefractive effects, which further diminish the possibility of perfect matching. It remains unclear whether high conversion efficiency can be achieved with detuned frequencies or in the presence of such parasitic effects.

In this Letter, we provide universal guidelines for optimizing cavity-based frequency conversion with continuous-wave pumping. 
We explicitly demonstrate the approach to maximize the conversion efficiency of second-harmonic generation (SHG) and sum-frequency-generation (SFG) in microresonators. Furthermore, we show in many cases that introducing a suitable amount of deviation from perfect matching can improve the performance of the devices. Finally, we show that the upper bound on conversion efficiency can be met even in the presence of parasitic processes, if certain design criteria are satisfied.

We model the microresonator-based $\chi^{(2)}$ processes [Fig. \ref{figScheme}(b)] with the coupled-mode equations \cite{Buryak_PR_2002,Sturman_JOSAB_2011,Ricciardi_PRA_2015},
\begin{align}
	&\frac{dA}{dt} = -\frac{\alpha_A}{2}A - i\Delta_A A + i\omega_A \kappa B^\ast C  + \sqrt{\frac{\theta_A}{t_R}}A_{\mathrm{in}},\label{SFG_1}\\
	&\frac{dB}{dt} = -\frac{\alpha_B}{2}B - i\Delta_B B + i\omega_B \kappa A^\ast C + \sqrt{\frac{\theta_B}{t_R}}B_{\mathrm{in}},\label{SFG_2}\\
	&\frac{dC}{dt} = -\frac{\alpha_C}{2}C - i(\delta_m+\Delta_A+\Delta_B) C + il\omega_C \kappa AB \notag\\&\quad\quad\quad+ \sqrt{\frac{\theta_C}{t_R}}C_{\mathrm{in}},\label{SFG_3}
\end{align}
where $A$, $B$, and $C$ are the intracavity field amplitudes of the $\omega_A$, $\omega_B$, and $\omega_C = \omega_A+\omega_B$ fields, respectively, $\alpha_A$, $\alpha_B$, and $\alpha_C$ are the respective loss rates, $\Delta_A$ and $\Delta_B$ are the detunings of the $\omega_A$ and $\omega_B$ fields from the corresponding cavity resonances, $\kappa$ is the nonlinear coefficient (Supplementary Material), $A_\mathrm{in}$, $B_\mathrm{in}$, and $C_\mathrm{in}$ are the input amplitudes for the $\omega_A$, $\omega_B$, and $\omega_C$ fields, respectively, $t_R$ is the roundtrip time, $l$ is a combinatoric coefficient which is 0.5 (1) in the case of $A$ and $B$ being degenerate (nondegenerate), $\theta_A$, $\theta_B$, and $\theta_C$ are the coupling rates of $\omega_A$, $\omega_B$, and $\omega_C$ fields, respectively, and $\delta_m = \Omega_C-\Omega_A-\Omega_B$ is the cavity resonance mismatch term determined by dispersion, where $\Omega_A$, $\Omega_B$, and $\Omega_C$ are the cavity resonance frequencies corresponding to the $A$, $B$, and $C$ fields, respectively. We define $\Delta_{A,B}=\Omega_{A,B}-\omega_{A,B}$ and the fields are in units of W$^{\frac{1}{2}}$. The output fields can be calculated as $X_\mathrm{out} = X_\mathrm{in} -\sqrt{\theta_Xt_R}X$, where $X \in \{A, B, C\}$. Under the mean-field assumption \cite{Stegeman_OQE_1996,Leo_PRA_2016}, phase mismatch of the nonlinear process is fully incorporated into the nonlinear coefficient $\kappa$ (see Supplementary Material). 

We first consider the SHG process in steady-state which corresponds to $A$ and $B$ fields being the degenerate pump field and $C$ being the second harmonic (SH) field with $C_\mathrm{in}=0$. The efficiency versus pump-power relation [see Fig \ref{figScheme}(c)] can be divided into 3 regimes for increasing pump powers: i) under-pumped, ii) saturation, and iii) over-pumped regimes, which corresponds to $\omega_A^2\kappa^2\mathcal{F}P_\mathrm{in}$ being much less than, comparable to, or much larger than $\alpha_C^2$, respectively, where $\mathcal{F}$ is the finesse of the pump resonance. In the under-pumped regime, where the SH field has negligible influence on the pump field, the power conversion efficiency can be expressed as,
\begin{align}
	\eta = \frac{64\omega_A^2\kappa^2\theta_A^2\theta_C }{t_R(4\Delta_A^2+\alpha_A^2)^2[\alpha_C^2+4(2\Delta_A+\delta_m)^2]}P_\mathrm{in},\label{CE_SS}
\end{align}
where $P_\mathrm{in}=|A_\mathrm{in}|^2$ is the pump power and the dependence of $\theta_C$ is due to output coupling. The maximum ratio of $\eta/P_\mathrm{in}$ is achieved with the perfect-matching condition, and, for fixed propagation losses, the highest efficiency is achieved when both the pump and SH fields are critically coupled.

However, a critically coupled cavity is not optimal for efficient frequency conversion in the saturation regime in which the pump field begins to experience depletion due to nonlinear interaction. The analytical expressions for the conversion efficiency in this regime are complicated and do not yield significant intuition. Instead, we derive the upper bound of the  efficiency, which can be shown to have the following simple form, 
\begin{align}
	\eta \le \frac{\theta_A\theta_C}{\alpha_A\alpha_C}.\label{SHG_Bound}
\end{align}
The conditions for meeting this bound are given by,
\begin{align}
	&\Delta_A = \frac{\alpha_A\delta_m}{\alpha_C-2\alpha_A},\label{SHG_Opt_1}\\
	&P_\mathrm{in} = \frac{\alpha_A\alpha_Ct_R}{\omega_A^2\kappa^2\theta_A}\left(\Delta_A^2+\frac{\alpha_A^2}{4}\right).\label{SHG_Opt_2}
\end{align}
Since two constraints must be satisfied to reach the maximum efficiency, two tunable parameters are required in practical applications. For a typical SHG device, these can be either the pump detuning $\Delta_A$, power $P_\mathrm{in}$, or the microresonator temperature ($\Delta_A$ and $\delta_m$ simultaneously). In many applications, such as for tunable coherent sources, the pump power and frequency are determined by the application and cannot be tuned to maximize the efficiency. We can introduce an additional tuning mechanism by using an auxiliary cavity [Fig \ref{figScheme}(a)] that can shift the SH resonance frequency without influencing the pump resonance \cite{Little_JLT_1997}. For example, an auxiliary cavity resonating near $\omega_C$ modifies the linear loss of the SH field by $\mu^2\alpha/(\alpha^2+\Delta^2)$, and modifies the resonance mismatch by $-\mu^2\Delta/(\alpha^2+\Delta^2)$, where $\mu$ is the coupling rate between the cavities, $\alpha$ is the loss rate of the auxiliary cavity, and $\Delta$ is the detuning of the auxiliary cavity from $\omega_C$ (Supplementary Material). The auxiliary cavity should be suitably detuned to provide resonance shift while avoiding excess losses. Alternatively, we can use a microresonator whose SH resonance linewidth is twice of that of the pump as explained further below in the discussion of the over-pumped regime.

Under the perfect matching condition, the conversion efficiency scales as $1/\sqrt[3]{P_\mathrm{in}}$ in the over-pumped regime. However, with adjustable pump detuning, the efficiency can be maximized to approximately the constant value,
\begin{align}
	\eta \approx \frac{8\theta_A\theta_C}{(2\alpha_A+\alpha_C)^2},\label{LS_Bound}
\end{align}
for a pump detuning $\Delta_A \approx \pm \frac{2\sqrt{2}\omega_A\kappa}{\alpha_C+2\alpha_A}\sqrt{\frac{\theta_A}{t_R}P_\mathrm{in}}$, which exceeds the efficiency with perfect matching. This asymptotic behavior explains experimental observations that showed flattening of conversion efficiency at high pump powers, which contradict the predictions of the simplified models \cite{Ilchenko_PRL_2004,Furst_PRL_2010,Lu_Optica_2019,Chen_arXiv_2021}. Since the optimal condition only involves detuning, simple temperature tuning is sufficient to meet this efficiency bound, which is ,in general, lower than the absolute bound. With an additional constraint $\alpha_C = 2\alpha_A$, the over-pumped efficiency bound is identical to the absolute bound of Eq. (\ref{SHG_Bound}). Nonlinear bistability has been observed in the saturation and over-pumped regimes \cite{Drummond_Misc_1980,Marte_PRA_1994,Trillo_OL_1996,Etrich_PRE_1997}, which can lead to instability. The subsequent self-pulsing and comb formation has been the focus of many studies. However, for efficient frequency conversion, these processes must be supressed. Following the methods used in \cite{Marte_PRA_1994,Trillo_OL_1996,Leo_PRA_2016}, we have analyzed both temporal and frequency stabilities of the SHG system, \textit{i.e.} the stability against temporal perturbations in Eqs. (\ref{SFG_1}-\ref{SFG_3}) and the possibility that additional frequency modes could be excited via modulation instabilities (Supplementary Material). Both the high-efficiency states in the saturation and overpumped regimes have large stable zones for the condition where the optimal detunings have opposite signs with respect to the group-velocity-dispersion (GVD) parameter ($\beta_2$). The stable zone can be further extended to include even higher input powers by increasing the ratio between the FSR difference an GVD at the pump frequency. As large FSR differences (\textit{i.e.} $>$ 1\% of the FSR) are generally expected due to the large frequency separation between the pump and SH modes, choosing $\Delta_A$ to have the opposite sign to $\beta_2$ (or vice versa) is generally sufficient to ensure stability of the high-efficiency states.

\begin{figure}
	\centering
	\includegraphics{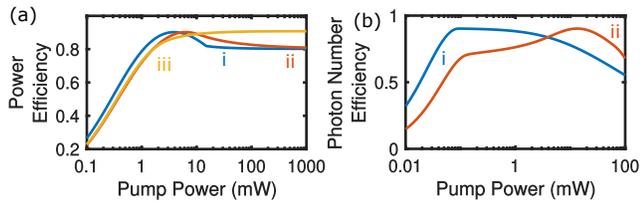}
	\caption{Numerical simulation for achieving efficient (a) SHG and (b) SFG in overcoupled PPLN microresonator. The pump detuning is optimized for each power level. In both figures, trace (i) corresponds to resonance matched cavities, and trace (ii) corresponds to resonance mismatched cavities. Trace (iii) in (a) corresponds to $\alpha_C=2\alpha_A$. (a) For (i) and (ii), $\alpha_A = \alpha_C =2\pi\times$5 GHz, and for (iii), $\alpha_A = \frac{\alpha_C}{2} =2\pi\times$4 GHz. For (i), $\delta_m$ = 0, and for (ii) and (iii), $\delta_m = 2\pi\times$2 GHz. (b) (i) represents matched resonances for SFG, \textit{i.e.}, $\delta_m$ = 0, and (ii) represents mismatched resonances with $\delta_m = 2\pi\times$3 GHz. $\alpha_A = 500$ MHz, $\alpha_B = \alpha_C =2\pi\times$5 GHz, and $\Delta_B = 0$.}
	\label{figHighEff}
\end{figure}

To illustrate our results, we simulate SHG in a periodically-poled lithium niobate (PPLN) microresonator. We assume a pump wavelength of 1.55 $\mu$m, a roundtrip time of 5 ps, and an intrinsic linewidth of 250 MHz for all resonances, which is well within the range of the current fabrication capabilities \cite{Zhang_Optica_2017}. We further assume that the $d_{33}$ nonlinearity is utilized, which corresponds to $\omega_A\kappa \approx$ 100 GHz/W$^{\frac{1}{2}}$. We numerically solve Eqs. (\ref{SFG_1}) and  (\ref{SFG_3}) in steady state and at each input pump power we optimize the detuning to achieve the highest conversion efficiency, as is typically done in experiments. As shown in Fig. \ref{figScheme}(c), trace (i) corresponds to a critically-coupled cavity with perfect matching, which has the highest efficiency at low pump powers. However, the maximum efficiency in the saturation (25\%) and over-pumped (22\%) regimes is the same as that of trace (ii), which has a 500-MHz resonance mismatch and is lower than that of trace (iii), which is both overcoupled and resonance mismatched. Without pump frequency tuning, resonance mismatch and over-pumping can significantly reduce the efficiency, as shown in traces (iv) and (v). To achieve an efficiency of > 90\%, the cavity should be highly overcoupled. For an intrinsic linewidth of 250 MHz, a loaded linewidth of 5 GHz is required regardless of the resonance-matching condition, as shown by traces (i) and (ii) in Fig \ref{figHighEff}(a). To demonstrate high conversion efficiency in the over-pumped regime, we choose a cavity linewidth of 4 GHz for the pump field and 8 GHz for the SH field as is shown by trace (iii), where a flat efficiency versus power curve is achieved for a large range of pump powers.

SFG is often used for quantum frequency conversion, where near unity efficiency is critical.  For such applications, the most relevant metric is the photon-number efficiency, which is defined as $\eta_\mathrm{ph} = \frac{\omega_B|C_\mathrm{out}|^2}{\omega_C|B_\mathrm{in}|^2}$. We show the analysis, where $A$, $B$, and $C$ correspond to a strong pump field and weak signal and idler fields, respectively. The efficiency bound can be shown to be,
\begin{align}
	\eta_\mathrm{ph} \le \frac{\theta_B\theta_C}{\alpha_B\alpha_C},\label{SFG_Bound}
\end{align}
for which the equality is met for,
\begin{align}
	&\Delta_A = (\frac{\alpha_C}{\alpha_B}-1)\Delta_B-\delta_m,\label{SFG_Opt_1}\\
	&P_\mathrm{in} = \frac{\alpha_Ct_R}{\alpha_B\kappa^2\omega_B\omega_C\theta_A}\left(\frac{\alpha_A^2}{4}+\Delta_A^2\right)\left(\frac{\alpha_B^2}{4}+\Delta_B^2\right),\label{SFG_Opt_2}
\end{align}
where $P_\mathrm{in}$ is the pump power. Similar to the SHG case, two tunable parameters are needed to satisfy the two constraints, which can be chosen from pump power, pump detuning, signal detuning, and microresonator temperature. Different from SHG, the bound does not involve the input power of the signal or the coupling condition of the pump field. Since only one strong field is present in SFG, the steady states are always stable. As shown in Fig \ref{figHighEff}(b), we numerically solve for Eqs. (\ref{SFG_1})-(\ref{SFG_3}) at different pump power levels while optimizing the pump detuning for each power. We assume both the pump and signal fields are in the telecom band. Efficient SFG can be achieved with strongly overcoupled signal and idler resonances that have loaded linewidths of 5 GHz. The cavity resonance mismatch are compensated by pump detuning at the cost of increased pump power, and the highest efficiency is only achieved at an optimal pump power level.

It can be shown that the optimal-condition relations Eqs. (\ref{SHG_Opt_1}, \ref{SHG_Opt_2}) and Eqs. (\ref{SFG_Opt_1}, \ref{SFG_Opt_2}) follow the same underlining principles. In addition to the conversion rate, the conversion efficiency of cavity-based processes depends on the coupling efficiency of the input field from the bus into the cavity, where the input field narrowly refers to the field that we wish to convert. In high $Q$ microresonators, the roundtrip coupling coefficient is typically < 1\%. Efficient coupling of the input field relies on the interference between the circulating and the incoming fields. Thus, an optimal conversion rate exists to maintain an optimal level of the circulating field. As detailed in the Supplementary Material, we can generalize the optimization rules for cavity-based frequency conversion as a nonlinear critical coupling of the input field: i) the conversion rate should be equal to the linear cavity-loss rate for the input field, and ii) the nonlinear phase shift should cancel the linear detuning for the input field. Importantly, these rules apply to nonlinear processes whether or not the conversion rate depends on the input field strength. Nonlinear critical coupling differ from phase matching in that the phase balancing is only required for the input field, which can be either pump or signal, rather than for both signal and idler fields as required by phase matching. In $\chi^{(2)}$-based processes, the nonlinear phase shift is due to a cascading of phase or resonance mismatched interactions, which yields an effective $\chi^{(3)}$ nonlinearity \cite{Belashenkov_Misc_1989,Stegeman_OQE_1996}.

The equation form of the optimal-condition relations allows us to easily incorporate parasitic effects. A large class of parasitic processes in $\chi^{(2)}$ microresonators can be represented as,
\begin{align}
	\begin{pmatrix}
		\Delta_A \\ \Delta_B \\ \delta_m
	\end{pmatrix}	= 
	\begin{pmatrix}
		\Delta_{A0} \\ \Delta_{B0} \\ \delta_{m0}
	\end{pmatrix} + \mathbf{M}
	\begin{pmatrix}
		|A|^2 \\ |B|^2 \\ |C|^2
	\end{pmatrix},\label{Paras}
\end{align}
where $\Delta_{A0}$ and $\Delta_{B0}$ correspond to the detunings from the linear cavity, $\delta_{m0}$ corresponds to the resonance mismatch of the linear cavity, in the absence of parasitic effects, and $\mathbf{M} = (M_{ij})_{3\times3}$ corresponds to the strengths of power-dependent resonance shifts. This model can represent photorefractive, thermorefractive, self-phase modulation, and cross-phase modulation effects. We substitute Eq. (\ref{Paras}) into Eq. (\ref{SHG_Opt_1}) to reach the modified condition for maximum SHG efficiency (Supplementary Material),
\begin{align}
	4G\Delta_A^2+(2\alpha_A-\alpha_C)\Delta_A+\alpha_A\delta_{m0}+G\alpha_A^2=0,\label{SHG_Paras_1}
\end{align}
where $G=\frac{M_{33}\alpha_A+M_{31}\alpha_C}{4\omega_A^2\kappa^2}$, and,
\begin{align}
	\Delta_A-\frac{M_{13}\alpha_A+M_{11}\alpha_C}{4\alpha_A\omega_A^2\kappa^2}(4\Delta_A^2+\alpha_A^2) = \Delta_{A0}.\label{SHG_Paras_2}
\end{align}
Equations (\ref{SHG_Paras_1}) and (\ref{SHG_Paras_2}) should be regarded as a single constraint relating the experimentally tunable parameters $\Delta_{A0}$ and $\delta_{m0}$. Importantly, to meet the efficiency bound, the resonance mismatch term $\delta_{m0}$ should be set within a certain range such that Eq. (\ref{SHG_Paras_1}) has real roots for $\Delta_A$. A similar constraint can be derived for SFG, where we assume the parasitic phase shifts are produced solely by the strong pump field, such that,
\begin{align}
	4H\Delta_B^2-\Delta_B+\Delta_{B0}+H\alpha_B^2 = 0,\label{SFG_Paras_1}
\end{align}
where $H=\frac{M_{21}\alpha_C}{4\alpha_B\omega_B\omega_C\kappa^2}$, and,
\begin{align}
	\frac{\alpha_C-\alpha_B}{\alpha_B}\Delta_B-\frac{(M_{11}+M_{31})\alpha_C}{\alpha_B\omega_B\omega_C\kappa^2}(\Delta_B^2+\frac{\alpha_B^2}{4}) = \Delta_{A0}+\delta_{m0}.\label{SFG_Paras_2}
\end{align}
These conditions indicate that with a suitable choice of initial input signal detuning, the SFG efficiency bound can also be met.

\begin{figure}
	\centering
	\includegraphics{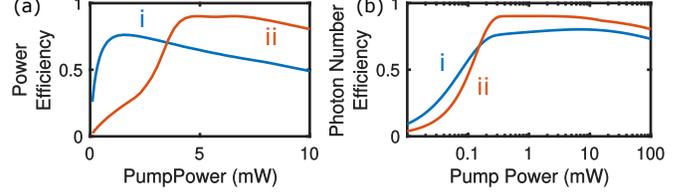}
	\caption{Numerical simulation of (a) SHG and (b) SFG efficiency in PPLN microresonator in the presence of the photorefractive effect.  Pump detunings are optimized for each power level. (a) (i) corresponds to matched resonances and (ii) corresponds to $\delta_{m0} = 2\pi\times$10 GHz initial resonance mismatch. $\alpha_A = \alpha_C =2\pi\times$5 GHz. (b) (i) corresponds to matched resonances and (ii) corresponds to $\Delta_{B0}=2\pi\times$(-2) GHz. $\alpha_B = \alpha_C =2\pi\times$5 GHz, $\delta_{m0} = 0$. To better show the consequences of the parasitic effect, we use $\omega_A\kappa=50$ GHz/(W)$^\frac{1}{2}$ in the SFG simulation.}
	\label{figParasitics}
\end{figure}

As an example, we consider PPLN microresonators with the photorefractive effect. The coefficients $M_{ij}$ vary substantially between different device structures \cite{Xu_arXiv_2020} and doping \cite{Buse_PM}. We estimate the values based on the measurements reported in \cite{Jiang_OL_2017}(Supplementary Material). We compare perfectly resonance-matched cavities [traces (i) in Fig. \ref{figParasitics}(a) and (b)] with cavities having suitably mismatched or detuned resonances [traces (ii) in Fig. \ref{figParasitics}(a) and (b)]. At each pump power level, we optimize the pump detuning to achieve the corresponding maximum conversion efficiency. As predicted by the analytical analysis, the resonance matched cavities cannot reach the upper bound of the conversion efficiency (90\%), which is remedied by introducing a suitable amount of resonance mismatch or signal detuning.

In conclusion, we have identified the efficiency upper bounds for $\chi^{(2)}$-microresonator-based frequency conversion, and the conditions for meeting such bounds. To compensate for fabrication errors and environmental variations, two tunable parameters are required to maximize the efficiency of individual devices, which corresponds a nonlinear-critical-coupling condition. Our analysis provides guidelines for the cavity designs for high-efficiency SHG with largely tunable input frequencies and powers, high-efficiency SFG with fixed-frequency pump and signal, and high-efficiency SHG and SFG with strong parasitic effects, \textit{etc}. Based on our analysis, near unity efficiency can be achieved in practical microresonators.
\\[6pt]
\textbf{Funding.}
\begingroup
\fontsize{8pt}{6pt}\selectfont
We acknowledge support from National Science Foundation (NSF-2040702).
\endgroup
\\[6pt]
\textbf{Disclosures.}
\begingroup
\fontsize{8pt}{6pt}\selectfont
The authors declare no conflicts of interest.
\endgroup
\\[6pt]
\textbf{Data availability.}
\begingroup
\fontsize{8pt}{6pt}\selectfont
Data underlying the results presented in this paper are not publicly available at this time but may be obtained from the authors upon reasonable request.
\endgroup
\\[6pt]
\textbf{Supplemental document.}
\begingroup
\fontsize{8pt}{6pt}\selectfont
See Supplement 1 for supporting content.
\endgroup

\newpage 
\widetext
\begin{center}
	\textbf{\large Theory of $\chi^{(2)}$-microresonator-based frequency conversion: supplementary material}
\end{center}
\setcounter{equation}{0}
\setcounter{figure}{0}
\setcounter{table}{0}
\setcounter{page}{1}
\setcounter{section}{0}
\makeatletter
\renewcommand{\theequation}{S\arabic{equation}}
\renewcommand{\thefigure}{S\arabic{figure}}
\renewcommand{\bibnumfmt}[1]{[S#1]}
\renewcommand{\citenumfont}[1]{S#1}

\section{Nonlinear coefficient}
In this section, we give the full expression of the nonlinear coefficient, which first appears in Eqs. (1-3). The nonlinear coefficient is defined as \cite{Sturman_JOSAB_2011},
\begin{align}
	\kappa = \frac{\chi^{(2)}L}{ct_R\sqrt{2n_An_Bn_C\epsilon_0cS}}K,
\end{align}
where $\chi^{(2)}$ is the bulk nonlinear coefficient, $L$ is the cavity length, $c$ is the speed of light, $n_A$, $n_B$ and $n_C$ are the effective refractive indices of the field $A$, $B$, and $C$, respectively, $\epsilon_0$ is the vacuum permittivity, $S$ is mode overlap area, and $K$ is a dimensionless factor accounting for potential reductions of nonlinearity such as periodic poling or phase mismatch \cite{Leo_PRA_2016}. A phase mismatch $\Delta\beta$ leads to $K=\mathrm{sinc}(\Delta\beta L/2)$. The overlap area is defined as,
\begin{align}
	S = \frac{\int|f_A(x,y)|^2dxdy\int|f_B(x,y)|^2dxdy\int|f_C(x,y)|^2dxdy}{|\int f_A(x,y)f_B(x,y)f_C^*(x,y)dxdy|^2},\label{S_Kappa}
\end{align}
where $f_A$, $f_B$, and $f_C$ are the mode profiles for the $A$, $B$, and $C$ fields, respectively. Equation (\ref{S_Kappa}) is a good approximate form for most waveguide strucutres. A more rigorous definition can be found in \cite{Yang_OL_2007}.

\section{Conversion efficiency of second-harmonic generation}
In this section, we provide the detailed derivations of conversion efficiency in second-harmonic generation (SHG) that lead to Eqs. (4-8) in the main text. In order to study the intracavity behavior, it is convenient to use normalized equations. We perform the substitutions $\tau = \frac{\alpha_A}{2}t$, $\alpha = \frac{\alpha_C}{\alpha_A}$, $\delta = \frac{2\Delta_A}{\alpha_A}$, $\sigma = \frac{2\delta_m}{\alpha_A}$, $a = \frac{2\omega_A\kappa}{\alpha_A}A$, $c = \frac{2\omega_A\kappa}{\alpha_A}C$, and $F = \frac{4\omega_A\kappa}{\alpha_A^2}\sqrt{\frac{\theta_A}{t_R}}A_\mathrm{in}$, which yields the normalized equations,
\begin{align}
	&\frac{da}{d\tau} = -a - i\delta a + ia^\ast c +F,\label{S_SHG_Field_1}\\
	&\frac{dc}{d\tau} = -\alpha c - i\sigma c - i2\delta c + ia^2.\label{S_SHG_Field_2}
\end{align}
In this paper, we focus on the steady state behavior, which can be analyzed by setting the time derivatives to 0. With some manipulations, we get two real-valued polynomial equations,
\begin{align}
	&|a|^6 - 2(\delta \tilde{\sigma}-\alpha)|a|^4 + (\delta^2+1)(\alpha^2+\tilde{\sigma}^2)|a|^2-(\alpha^2+\tilde{\sigma}^2)|F|^2=0,\label{S_SHG_Norm_1}\\
	&|c|^2 = \frac{|a|^4}{\alpha^2+\tilde{\sigma}^2},\label{S_SHG_Norm_2}
\end{align}
where $\tilde{\sigma} = 2\delta+\sigma$. For the under-pumped regime, we can ignore the $|a|^6$ and $|a|^4$ terms in Eq. (\ref{S_SHG_Norm_1}). It is straightforward to find that,
\begin{align}
	\frac{|c|^2}{|F|^2} = \frac{|F|^2}{(\delta^2+1)^2[\alpha^2+(2\delta+\sigma)^2]}.\label{S_SHG_SS}
\end{align}
The input-output relation shown in the main text reads, 
\begin{align}
	X_\mathrm{out} = X_\mathrm{in} -\sqrt{\theta_Xt_R}X,\label{S_Output}
\end{align}
where $X \in \{A, B, C\}$. Using Eq. (\ref{S_Output}) and the normalization conditions in this section, Eq. (\ref{S_SHG_SS}) can be converted to Eq. (4) in the main text. 

The upper bound of conversion efficiency in all regimes can be found using two inequalities,
\begin{align}
	&|a|^2 + \frac{(\delta^2+1)(\alpha^2+\tilde{\sigma}^2)}{|a|^2} \ge 2\sqrt{(\delta^2+1)(\alpha^2+\tilde{\sigma}^2)},\label{S_Ineq1}\\
	&\sqrt{(\delta^2+1)(\alpha^2+\tilde{\sigma}^2)} \ge \delta \tilde{\sigma} + \alpha.
\end{align}
It then follows,
\begin{align}
	\frac{|c|^2}{|F|^2} = \frac{1}{|\alpha|^2+\frac{(\delta^2+1)(\alpha^2+\tilde{\sigma}^2)}{|\alpha|^2}-2(\delta\tilde{\sigma}-\alpha)} \le \frac{1}{4\alpha}, \label{S_SHG_Norm_Bound}
\end{align}
where the equality is satisfied when,
\begin{align}
	&\delta = \frac{\sigma}{\alpha-2},\label{S_SHG_Opt_1}\\
	&|F|^2 = 4\alpha(\delta^2+1).\label{S_SHG_Opt_2}
\end{align}
Using Eq. (\ref{S_Output}) and the normalization conditions in this section, Eqs. (\ref{S_SHG_Norm_Bound}-\ref{S_SHG_Opt_2}) can be converted to Eqs. (5-7) in the main text. 

In the over-pumped regime, for fixed detuning, we can drop the $|a|^4$ and $|a|^2$ terms in Eq. (\ref{S_SHG_Norm_1}), which yields,
\begin{align}
	\frac{|c|^2}{|F|^2} = \frac{1}{\sqrt[3]{(\alpha^2+\tilde{\sigma}^2)|F|^2}}.\label{S_SHG_OP_Fixed}
\end{align}
Equation (\ref{S_SHG_OP_Fixed}) indicates that the efficiency scales as $1/\sqrt[3]{P_\mathrm{in}}$, where $P_\mathrm{in}$ is the input pump power. However, the scaling can be improved if we consider $\delta$ to be adjustable. Particularly, if $\delta$ is comparable to $|F|$, the assumption of $|a|^4$ and $|a|^2$ terms being negligible becomes invalid. To treat this, we combine Eqs. (\ref{S_SHG_Norm_1}), (\ref{S_SHG_Norm_2}), and (\ref{S_Ineq1}) to get,
\begin{align}
	\frac{|c|^2}{|F|^2} \le \frac{1}{2\delta^2(\sqrt{(1+\frac{1}{\delta^2})(4+\frac{4\sigma}{\delta}+\frac{\alpha^2+\sigma^2}{\delta^2})}-2\frac{\sigma}{\delta})+2\alpha}.
\end{align}
By expanding the denominator to the second order of the small parameter $\frac{1}{\delta}$, we get a simplified expression,
\begin{align}
	\frac{|c|^2}{|F|^2} \le \frac{2}{(\alpha+2)^2+O(\frac{1}{\delta})}.\label{S_SHG_Op}
\end{align}
This bound can be approached with,
\begin{align}
	\delta \approx \pm\frac{\sqrt{2}|F|}{\alpha+2}.\label{S_SHG_Over_Opt}
\end{align}
Note, when $\alpha = 2$, the large-signal bound is identical to the absolute bound Eq. (\ref{S_SHG_Norm_Bound}). This agrees with the optimal condition Eq. (\ref{S_SHG_Opt_1}) since it requires large $\delta$ for $\alpha \approx 2$, which is also the condition of the over-pumped regime. Using Eq. (\ref{S_Output}) and the normalization conditions in this section, Eq. (\ref{S_SHG_Op}) can be converted to Eq. (8) in the main text.

\section{Conversion efficiency of sum-frequency generation}
In this section, we provide derivations of conversion efficiency in sum-frequency generation (SFG) that lead to Eqs. (9-11) in the main text. We also use normalized equations to study the cavity-based SFG. The normalization of these equations follows $\tau = \frac{\alpha_A t}{2}$, $\alpha_b = \frac{\alpha_B}{\alpha_A}$, $\alpha_c = \frac{\alpha_C}{\alpha_A}$, $\delta_a = \frac{2\Delta_A}{\alpha_A}$, $\delta_b = \frac{2\Delta_B}{\alpha_A}$, $\sigma = \frac{2\delta_m}{\alpha_A}$, $a = \frac{2\kappa\sqrt{\omega_B\omega_C}}{\alpha_A}A$, $b = \frac{2\kappa\sqrt{\omega_A\omega_C}}{\alpha_A}B$, $c = \frac{2\kappa\sqrt{\omega_A\omega_B}}{\alpha_A}C$, $F_a = \frac{4\kappa\sqrt{\omega_B\omega_C\theta_A}}{\alpha_A^2\sqrt{t_R}}A_\mathrm{in}$, and $F_b = \frac{4\kappa\sqrt{\omega_A\omega_C\theta_B}}{\alpha_A^2\sqrt{t_R}}B_\mathrm{in}$, which yields,
\begin{align}
	&\frac{da}{d\tau} = -a-i\delta_a a + ib^\ast c + F_a,\label{S_SFG_Norm_1}\\
	&\frac{db}{d\tau} = -\alpha_b b - i\delta_b b + ia^\ast c + F_b,\label{S_SFG_Norm_2}\\
	&\frac{dc}{d\tau} = -\alpha_c c - i\sigma c - i(\delta_a+\delta_b)c + ia b. \label{S_SFG_Norm_3}
\end{align}
We focus on optimizing the conversion efficiency from signal $F_b$ to idler $c$ in steady state. We can find real valued equations connecting the two fields as,
\begin{align}
	\frac{|c|^2}{|F_b|^2} = \frac{1}{|a|^2+2(\alpha_b\alpha_c-\delta_b \tilde{\sigma}) + \frac{(\alpha_b^2+\delta_b^2)(\alpha_c^2+\tilde{\sigma}^2)}{|a|^2}},
\end{align}
where $\tilde{\sigma} = \sigma+\delta_a+\delta_b$. Similar to the SHG analysis, the upper bound of this value can be derived using two inequalities,
\begin{align}
	&|a|^2+\frac{(\alpha_b^2+\delta_b^2)(\alpha_c^2+\tilde{\sigma}^2)}{|a|^2} \ge 2\sqrt{(\alpha_b^2+\delta_b^2)(\alpha_c^2+\tilde{\sigma}^2)},\\
	&\sqrt{(\alpha_b^2+\delta_b^2)(\alpha_c^2+\tilde{\sigma}^2)} \ge (\delta_b \tilde{\sigma}+\alpha_b\alpha_c).
\end{align}
The corresponding bound is found as,
\begin{align}
	\frac{|c|^2}{|F_b|^2} \le \frac{1}{4\alpha_b\alpha_c},\label{S_SFG_Norm_Max}
\end{align}
where the equality is reached with,
\begin{align}
	&\delta_a = \left(\frac{\alpha_c}{\alpha_b}-1\right)\delta_b-\sigma,\label{S_SFG_Opt_1}\\
	&|F_a|^2 = \frac{\alpha_c}{\alpha_b}\left(\alpha_b^2+\delta_b^2\right)\left[(1+\delta_a^2)+\frac{(\alpha_c-\delta_a \tilde{\sigma})|F_b|^2}{2\alpha_c^2(\alpha_b^2+\delta_b^2)}+\frac{|F_b|^4}{16\alpha_b^2\alpha_c^2(\alpha_b^2+\delta_b^2)}\right].\label{S_SFG_Opt_2}
\end{align}
For small $|F_b|$, Eq. (\ref{S_SFG_Opt_2}) can be simplified into,
\begin{align}
	|F_a|^2 = \frac{\alpha_c}{\alpha_b}(\alpha_b^2+\delta_b^2)(1+\delta_a^2).\label{S_SFG_Opt_Simp}
\end{align}
Using Eq. (\ref{S_Output}) and the normalization conditions in this section, Eqs. (\ref{S_SFG_Norm_Max}), (\ref{S_SFG_Opt_1}), and (\ref{S_SFG_Opt_Simp}) can be converted to Eqs. (9-11) in the main text.

\section{Physical interpretation of the optimal condition}
In this section, we analyze the optimal-condition relations for SHG and SFG and justify the nonlinear-critical-coupling condition proposed in the main text. At steady state, the pump field of the SHG process satisfies,
\begin{align}
	-a-i\delta a -\frac{|a|^2}{\alpha+i\sigma+i2\delta}a + F = \frac{da}{d\tau} = 0,\label{S_SHG_Field_3}
\end{align}
where all variables are defined in section 2. The left-hand-side terms with real coefficients corresponds to losses and imaginary coefficients corresponds to detuning. The nonlinear (third) term has two effects, which are intensity dependent loss and detuning. The intensity dependent loss is an alternative expression of conversion rate, and the intensity dependent detuning is the result of an effective self-phase modulation (SPM) process from cascaded $\chi^{(2)}$ processes \cite{Belashenkov_Misc_1989,Stegeman_OQE_1996}. To understand the optimal condition, we first substitute the optimal condition [Eqs. (\ref{S_SHG_Opt_1}) and (\ref{S_SHG_Opt_2})] into Eqs. (\ref{S_SHG_Norm_1}) and (\ref{S_SHG_Norm_2}), which yields,
\begin{align}
	&|a|^2 = \alpha (\delta^2+1),\label{S_SHG_Opt_3}\\
	&|c|^2 = \delta^2+1.\label{S_SHG_Opt_4}
\end{align} 
Combining Eq. (\ref{S_SHG_Opt_3}) with Eq. (\ref{S_SHG_Opt_1}) and (\ref{S_SHG_Field_1}), the nonlinear term can be rewritten as,
\begin{align}
	-\frac{|a|^2}{\alpha + i\sigma+i2\delta}a = -a + i\delta a,\label{S_SHG_Critical_1}
\end{align}
which shows that the nonlinear loss (conversion rate) matches the linear loss and the nonlinear detuning cancels the linear detuning. This condition is analogous to a signal being critically coupled to a linear cavity, which requires the coupling rate being equal to the intrinsic loss rate of the cavity and the frequency of the signal being equal to the cavity resonance frequency. Thus we refer to it as nonlinear critical coupling. Equation (\ref{S_SHG_Critical_1}) is equivalent to Eqs. (\ref{S_SHG_Opt_1}) and (\ref{S_SHG_Opt_2}), which are the normalized versions of Eqs. (6) and (7), in defining the optimal-condition relations.

As shown in the main text, when $\alpha=2$, the highest efficiency can be reached in the over-pumped regime. With large detuning, the nonlinear term in Eq. (\ref{S_SHG_Field_3}) becomes,
\begin{align}
	-\frac{|a|^2}{\alpha + i\sigma+i2\delta}a \approx -\frac{\alpha|a|^2}{4\delta^2}a + i\frac{|a|^2}{2\delta} a.
\end{align}
When the nonlinear detuning cancels the linear detuning, the nonlinear loss coefficient is further reduced to $-\alpha/2$. When $\alpha = 2$, this rate is matched with the linear loss, fulfilling the nonlinear critical-coupling condition.

In the case of SFG, the weak signal at the steady state satisfies,
\begin{align}
	-\alpha_b b - i\delta_b b - \frac{|a|^2}{\alpha_c+i(\delta_a+\delta_b+\sigma)}b + F_b = \frac{db}{d\tau} = 0.\label{S_SFG_Interp}
\end{align}
Additionally, when the optimal conditions [Eqs. (\ref{S_SFG_Opt_1}) and (\ref{S_SFG_Opt_2})] are satisfied, we have,
\begin{align}
	|a|^2 = \frac{\alpha_c}{\alpha_b}(\alpha_b^2+\delta_b^2). \label{S_SFG_Opt_3}
\end{align}
Combining this with Eq. (\ref{S_SFG_Opt_1}), we can rewrite the nonlinear term (third term on the left hand side) of Eq. (\ref{S_SFG_Interp}) into,
\begin{align}
	-\frac{|a|^2}{\alpha_c+i(\delta_a+\delta_b+\sigma)}b = -\alpha_b b + i\delta_b b,\label{S_SFG_Critical}
\end{align}
which shows that the nonlinear loss equals linear loss and nonlinear detuning cancels linear detuning, which corresponds to the nonlinear critical-coupling condition. Equation (\ref{S_SFG_Critical}) is equivalent to Eqs. (\ref{S_SFG_Opt_1}) and (\ref{S_SFG_Opt_Simp}), which are normalized versions of Eqs. (10) and (11), in defining the optimal-condition relations.

\section{Analysis of coupled cavities}
\begin{figure}
	\centering
	\includegraphics{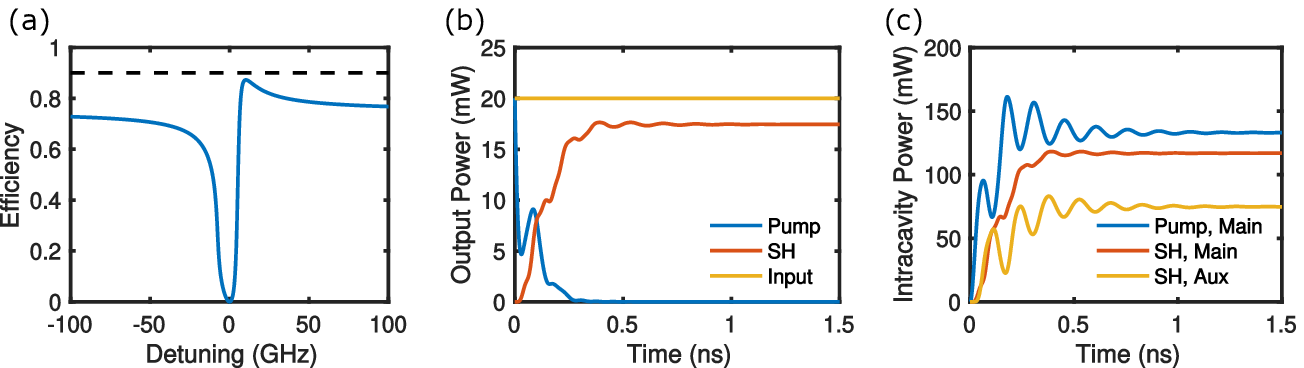}
	\caption{Simulations of SHG in coupled-cavity geometry. $\alpha_A=\alpha_C=2\pi\times$5 GHz, $\Delta_A = 2\pi\times$5.1 GHz, $\delta_m = 2\pi\times$2 GHz, $\mu = 2\pi\times$8 GHz. (a) Conversion efficiency as a function of auxiliary cavity detuning. (b) Temporal evolution of intracavity powers at $\Delta_D = 2\pi\times$10 GHz. (c) Temporal evolution of the output powers corresponding to (b).}
	\label{figAux}
\end{figure}

In this section, we analyze the performance of an auxiliary cavity [Fig. 1(a)] as a way to improve conversion efficiency. This approach can be used when the pump frequency and power are not adjustable. We assume a small portion of the idler field is coupled into an auxiliary resonance. This field is denoted as $D$. The governing equations for the idler field in the main cavity ($C$) and in the auxiliary cavity ($D$) can be written as \cite{Little_JLT_1997},
\begin{align}
	&\frac{dC}{dt} = -\frac{\alpha_C}{2}C - i\delta_m C - i(\Delta_A+\Delta_B) C + il\omega_C \kappa AB + i\mu D,\label{S_Coupled_1}\\
	&\frac{dD}{dt} = -\frac{\alpha_D}{2}D - i\Delta_D D + i\mu C,\label{S_Coupled_2}
\end{align}
where $\alpha_D$ is the power dissipation rate of the auxiliary cavity, $\Delta_D$ is the detuning of the auxiliary cavity from $\omega_C$, and $\mu$ is the coupling rate between the main cavity and the auxiliary cavity. At the steady state, the term due to the auxiliary cavity in Eq (\ref{S_Coupled_1}) (last term on the right hand side) can be expressed as,
\begin{align}
	i\mu D = -\frac{2\mu^2\alpha_D}{\alpha_D^2+4\Delta_D^2}C +i\frac{4\mu^2\Delta_D}{\alpha_D^2+4\Delta_D^2}C,
\end{align}
where the first term on the right hand side represents additional losses and the second term represents the shift of cavity resonance. As the loss scales as $1/\Delta_D^2$ and the resonance shift (dispersion) scales as $1/\Delta_D$, the auxiliary cavity should be configured to be suitably detuned from the idler frequency to reduce the coupling induced losses.

As an example, we use the cavity parameters of trace (ii) in Fig. 1(a) and aim to improve the efficiency at 20 mW pump power. We assume temperature tuning is applied to set the pump detuning such that Eq. (7) is satisfied. We further assume that the resonance mismatch after temperature tuning to be $\delta_m = 2\pi\times 2$ GHz. The auxiliary cavity is assumed to have a linewidth of 250 MHz, and the coupling rate between the cavities is $2\pi\times$8 GHz. Fig. \ref{figAux}(a) shows the conversion efficiency as a function of the auxiliary cavity detuning. As shown in Fig. \ref{figAux}(a), the introduction of the auxiliary cavity slightly reduces the maximum efficiency compared to the upper bound derived for the main cavity (dashed line). However, it is significantly higher than what is achievable with only pump-detuning optimization at 20 mW [Fig. 1(a)]. Fig. \ref{figAux}(b) and \ref{figAux}(c) show the temporal evolution of the introcavity and output power, respectively, at the maximum efficiency point. Even though a significant amount of power is present in the auxiliary cavity, the additional loss it introduces is small because of the low loss rate of the auxiliary cavity.

\section{Optimal conditions at the presence of parasitic effects}
In this section, we show that the optimal-condition relation derived in this paper can easily include parasitic effects. The results in this section directly lead to Eqs. (13-16) in the main text. To derive Eqs. (13) and (14) in the main text, we first rewrite Eqs. (\ref{S_SHG_Opt_3}) and (\ref{S_SHG_Opt_4}) in physical parameters,
\begin{align}
	&|A|^2 = \frac{\alpha_C}{4\omega_A^2\kappa^2\alpha_A}(4\Delta_A^2+\alpha_A^2),\label{S_SHG_Opt_Phys_A}\\
	&|C|^2 = \frac{1}{4\omega_A^2\kappa^2}(4\Delta_A^2+\alpha_A^2).\label{S_SHG_Opt_Phys_C}
\end{align}
Following Eq. (12), the parasitic effect can be represented as,
\begin{align}
	&\Delta_A = \Delta_{A0} + M_{11}|A|^2+M_{13}|C|^2,\label{S_SHG_Paras_1}\\
	&\delta_m = \delta_{m0} + M_{31}|A|^2+M_{33}|C|^2.\label{S_SHG_Paras_2}
\end{align}
Equation (13) can be derived by combining Eqs. (6), (\ref{S_SHG_Opt_Phys_A}), (\ref{S_SHG_Opt_Phys_C}), and (\ref{S_SHG_Paras_2}). Equation (14) can be derived by combining Eqs. (\ref{S_SHG_Opt_Phys_A}-\ref{S_SHG_Paras_1}).

To derive Eqs. (15-16) in the main text, we first rewrite Eq. (\ref{S_SFG_Opt_3}) in physical parameters,
\begin{align}
	|A|^2 = \frac{\alpha_C}{4\kappa^2\omega_B\omega_C\alpha_B}(4\Delta_B^2+\alpha_B^2).\label{S_SFG_Opt_Phys_A}
\end{align}
We assume the parasitic effect is solely due to the strong pump, thus Eq. (12) can be rewritten as,
\begin{align}
	&\Delta_A = \Delta_{A0} + M_{11}|A|^2,\label{S_SFG_Paras_1}\\
	&\Delta_B = \Delta_{B0} + M_{21}|A|^2,\label{S_SFG_Paras_2}\\
	&\delta_m = \delta_{m0} + M_{31}|A|^2.\label{S_SFG_Paras_3}
\end{align}
Equation (15) can be derived by combining Eqs. (\ref{S_SFG_Opt_Phys_A}) and (\ref{S_SFG_Paras_2}). Equation (16) can be derived by combining Eqs. (10), (\ref{S_SFG_Paras_1}), and(\ref{S_SFG_Paras_3}).

\section{Estimation of the M matrix}
In this section, we estimate the value of the $\textbf{M}$ matrix that first appears in Eq. (12). We estimate the $\textbf{M}$ matrix for the photorefractive effect based on the measurements in \cite{Jiang_OL_2017}. As the strengths and dispersive behaviors of the photorefractive effect depends heavily on the material preparation and fabrication, we do not seek to perform an accurate modeling of a particular device. Instead, we perform an order-of-magnitude estimation to demonstrate the principles of our optimization criteria. We assume that the optical fields induce changes in the refractive index that are proportional to the input powers. Importantly, we assume that the index change is constant across all wavelength regimes, \textit{i.e.}, the change is dispersionless, but its magnitude depends on the wavelengths of the electric fields. The detuning corresponding to the change of refractive index can be calculated as,
\begin{align}
	\Delta\Omega = \frac{\Omega}{n}\Delta n,\label{S_Index_Change}
\end{align}
where $\Omega$ ($\Delta\Omega$) is the (change of) resonance frequency, and $n$ ($\Delta n$) is the (change of) refractive index. We use the refractive index of bulk lithium niobate instead of the effective index for a particular waveguide structure. Based on \cite{Jiang_OL_2017}, we estimate a field near 1540 nm induces an index change of 6$\times 10^{-4}$/W and a field near 770 nm induces an index change of 0.07 /W, where, as the coefficients are intensity-dependent, we have scaled the mode volume to better match the $\kappa$ value used in our simulation. In our example, we assume both the pump and signal fields are in the telecom band and the idler field is near 770 nm. Using Eq. (\ref{S_Index_Change}), we can find $M_{11} = M_{12} = M_{21} = M_{22} = 2\pi\times$0.05 GHz/mW. $M_{13} = M_{23} = 2\pi\times $6 GH/mW. The $M_{3i}$ terms can be found according to $\Delta\Omega_C-\Delta\Omega_A-\Delta\Omega_B = (\frac{\Omega_C}{n(\Omega_C)}-\frac{\Omega_A}{n(\Omega_A)}-\frac{\Omega_B}{n(\Omega_B)})\Delta n$, which yields $M_{31} = M_{32} = 2\pi\times $(-0.002) GHz/mW, and $M_{33} = 2\pi\times $(-0.3) GHz/mW.

\section{Temporal stability of the high-efficiency states for SHG}
In this section, we prove the dynamical stability of the SHG states corresponding to conditions Eqs (5-7) (saturation regime) and Eq. (8) (over-pumped regime), respectively. The SFG stability is trivial since it only involves one strong pump that is not affected by the nonlinear process, thus it will not be discussed here. 

We analyze the stability of the high-efficiency states with normalized equations. We convert the complex-valued Eqs. (\ref{S_SHG_Field_1}) and (\ref{S_SHG_Field_2}) into real-valued ones with the substitution $a = x+iy$ and $c = u+iv$, which yields,
\begin{align}
	\frac{d}{d\tau}\begin{pmatrix}x\\y\\u\\v\end{pmatrix} = \textbf{f}(x,y,u,v) = \begin{pmatrix}
		-x+\delta y +yu-xv+F\\
		-y-\delta x +xu+yv\\
		-\alpha u + (\sigma+2\delta)v-2xy\\
		-\alpha v-(\sigma+2\delta)u+x^2-y^2
	\end{pmatrix}.
\end{align}
The Jacobian matrix of $\textbf{f}$ is given by,
\begin{align}
	\textbf{J}(x,y,u,v) = \frac{\partial\textbf{f}(x,y,u,v)}{\partial (x,y,u,v)} = \begin{pmatrix}
		-v-1 & \delta +u & y & -x \\
		u-\delta  & v-1 & x & y \\
		-2 y & -2 x & -\alpha  & \sigma+2\delta \\
		2 x & -2 y & -\sigma -2\delta  & -\alpha 
	\end{pmatrix}.\label{S_SHG_Jacobian}
\end{align} 
A steady state $(x_0,y_0,u_0,v_0)$ is stable if and only if all complex eigenvalues of $\textbf{J}(x_0,y_0,u_0,v_0)$ have nonpositive real parts. In general, calculating the eigenvalues of $\textbf{J}$ requires solving a quartic equation, which is difficult to study analytically. Instead, we prove the stability using the following three steps,
\begin{enumerate}
	\item[i)]{Each branch of eigenvalues is a continuous function of $\alpha$ and $\delta$.}
	\item[ii)]{An eigenvalue of $\textbf{J}$ cannot be a pure imaginary number.}
	\item[iii)]{For chosen values of $\alpha$ and $\delta$, the real parts of all eigenvalues are negative.}
\end{enumerate}
Step (i) is trivial since polynomial equations are analytic. The ``branches'' of solutions can also be rigorously defined by the root formula even though it is complicated for quartic equations. Step (ii) is to prove the trajectories of the eigenvalues cannot cross the imaginary axis of the complex plane, \text{i.e.}, the real part of a branch of eigenvalues is either always positive or always negative for all $\alpha$ and $\delta$. Thus, in step (iii), verifying the sign of the real part of one eigenvalue is sufficient to determine if the whole branch is located in the 2, 3 or 1, 4 quadrants of the complex plane.

\subsection{Saturation regime}
The high-efficiency state in the saturation regime corresponds to,
\begin{align}
	&x_0 = 	\sqrt{\alpha(\delta^2+1)},\\
	&y_0 = 0,\\
	&u_0 = \delta,\\
	&v_0 = 1.
\end{align}
Combining these with Eqs. (\ref{S_SHG_Opt_1}) and (\ref{S_SHG_Jacobian}), the Jacobian matrix can be reduced to,
\begin{align}
	\textbf{J}(x_0,y_0,u_0,v_0) = \left(
	\begin{array}{cccc}
		-2 & 2 \delta  & 0 & -\sqrt{\alpha  \left(\delta ^2+1\right)} \\
		0 & 0 & \sqrt{\alpha  \left(\delta ^2+1\right)} & 0 \\
		0 & -2 \sqrt{\alpha  \left(\delta ^2+1\right)} & -\alpha  & \alpha  \delta  \\
		2 \sqrt{\alpha  \left(\delta ^2+1\right)} & 0 & -\alpha  \delta  & -\alpha  \\
	\end{array}
	\right),
\end{align}
with the corresponding quartic equation for eigenvalue $\lambda$ being,
\begin{align}
	\lambda^4+2 (\alpha +1) \lambda^3+\alpha \left(\alpha \delta ^2+\alpha +4 \delta ^2+8\right) \lambda^2 +2 \alpha  (3 \alpha +2) \left(\delta^2+1\right) \lambda +8 \alpha ^2 \left(\delta ^2+1\right)=0. \label{S_Quartic_1}
\end{align}
To prove step (ii), we first assume $\lambda = iz$, where $z$ is a real number, is a solution of Eq. (\ref{S_Quartic_1}), which yields two equations,
\begin{align}
	&z^4-\alpha \left(\alpha \delta ^2+\alpha +4 \delta ^2+8\right)z^2+8 \alpha ^2 \left(\delta ^2+1\right)=0,\label{S_Stepii_1}\\
	&-2 (\alpha +1)z^3+2 \alpha  (3 \alpha +2) \left(\delta^2+1\right)z=0.\label{S_Stepii_2}
\end{align}
Since $\alpha>0$, $z=0$ is not a solution. Thus, we can substitute Eq. (\ref{S_Stepii_2}) into Eq. (\ref{S_Stepii_1}), which yields,
\begin{align}
	\left(3 \alpha ^3+8 \alpha ^2+10 \alpha +4\right) \delta ^4+2 \left(3 \alpha ^3+10 \alpha ^2+12 \alpha +4\right) \delta^2+ \left(3 \alpha ^3+12 \alpha ^2+14 \alpha+4\right)=0
\end{align}
Since $\alpha>0$, this equation cannot be satisfied. Thus we have proven step (ii). For $\delta=0$, Eq. (\ref{S_Quartic_1}) can be reduced to,
\begin{align}
	(\lambda^2+\alpha \lambda+2\alpha)(\lambda^2+\alpha \lambda+2\lambda+4\alpha) = 0
\end{align}
It is easy to show that if all coefficients of a quadratic equation are positive, the real parts of its roots are negative, which can be shown using the root formula for quadratic equations. This completes step (iii), and it follows that the high-efficiency state in the saturation regime is stable.

\subsection{Over-pumped regime}
The high-efficiency state in the over-pumped regime corresponds to,
\begin{align}
	&x_1 = \sqrt{\frac{\alpha^2+4\delta^2}{2}},\\
	&y_1 = 0,\\
	&u_1 = \delta,\\
	&v_1 = \frac{\alpha}{2}.
\end{align}
We assume $\delta = \sqrt{\frac{2F^2}{(\alpha+2)^2}-\frac{\alpha^2}{4}} \gg \sigma$, which yield,
\begin{align}
	\textbf{J}(x_1,y_1,u_1,v_1) = \left(
	\begin{array}{cccc}
		-\frac{\alpha }{2}-1 & 2 \delta  & 0 & -\sqrt{\frac{\alpha ^2+4 \delta ^2}{2}}\\
		0 & \frac{\alpha -2}{2} & \sqrt{\frac{\alpha ^2+4 \delta ^2}{2}} & 0 \\
		0 & -\sqrt{2\alpha ^2+8 \delta ^2} & -\alpha  & 2 \delta  \\
		\sqrt{2\alpha ^2+8 \delta ^2} & 0 & -2 \delta  & -\alpha  \\
	\end{array}
	\right).
\end{align}
The correponding quartic equation of eigenvalue $\lambda$ is,
\begin{align}
	\lambda^4+2 (\alpha +1) \lambda^3+ \left(\frac{11 \alpha ^2}{4}+4 \alpha +12 \delta^2+1\right)\lambda^2+\frac{1}{2} (\alpha +2) \left(3 \alpha ^2+2 \alpha +16 \delta ^2\right)\lambda&\notag\\+\frac{1}{4} \left(3 \alpha ^2+8 \alpha +4\right) \left(\alpha ^2+4 \delta^2\right)&=0.\label{S_Quartic_2}
\end{align}
If $\lambda=iz$, where $z$ is a real value, is a solution, we get,
\begin{align}
	&z^4-\left(\frac{11 \alpha ^2}{4}+4 \alpha +12 \delta^2+1\right)z^2+\frac{1}{4} \left(3 \alpha ^2+8 \alpha +4\right) \left(\alpha ^2+4 \delta^2\right)=0,\label{S_Stepii_3}\\
	&-2 (\alpha +1)z^3+\frac{1}{2} (\alpha +2) \left(3 \alpha ^2+2 \alpha +16 \delta ^2\right)z=0.\label{S_Stepii_4}
\end{align}
Since $\alpha>0$, we have $z \neq 0$. We substitute Eq. (\ref{S_Stepii_4}) into Eq. (\ref{S_Stepii_3}), which yields,
\begin{align}
	-\frac{3 \alpha ^6}{4}-\frac{65 \alpha ^5}{16}-\left(11 \delta^2+\frac{35}{4}\right)\alpha ^4 -\left(40 \delta ^2+\frac{19}{2}\right)\alpha ^3 -\left(32 \delta ^4+47 \delta ^2+5\right)\alpha ^2&\notag\\-\left(80 \delta ^4+24 \delta^2+1\right)\alpha  -4 \left(8 \delta ^4+\delta ^2\right)&=0
\end{align}
Since $\alpha>0$, this equation cannot be satisfied, which proves step (ii). Next we let $\delta=0$ in Eq. (\ref{S_Quartic_2}), which yields,
\begin{align}
	\left(\lambda^2+\frac{3\alpha+2}{2}\lambda+\frac{3\alpha^2+2\alpha}{2}\right)\left(\lambda^2+\frac{\alpha+2}{2}\lambda+\frac{\alpha^2+2\alpha}{2}\right)=0
\end{align}
Since $\alpha>0$, the real parts of the roots are all negative. Thus we have proven step (iii), which shows the high-efficiency steady state in the over-pumped regime is stable.

\section{Temporal evolution under the optimal conditions}
\begin{figure}
	\centering
	\includegraphics{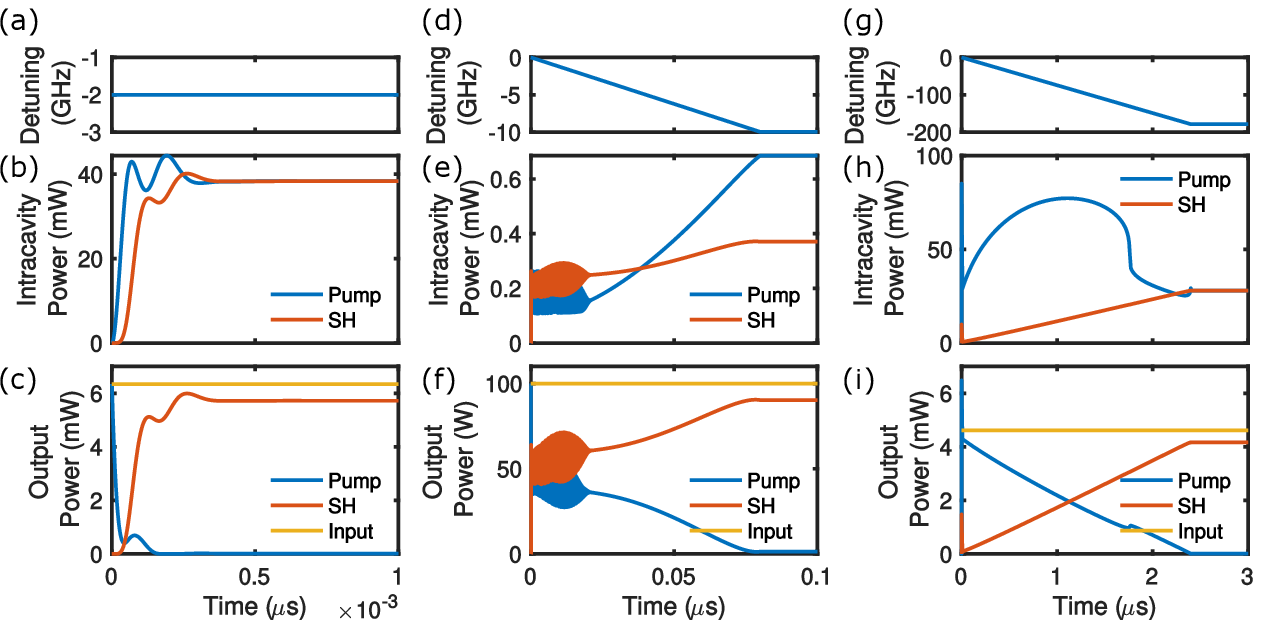}
	\caption{Temporal evolution of SHG for an intrinsic linewidth of $2\pi\times$250 MHz. (a), (d), and (g) show the tuning procedure. (b), (e), and (h) show the intracavity dynamics. (c), (f) and (i) show the evolution of the output fields. (a-c) High-efficiency state in the saturation regime. $\alpha_A=\alpha_C=2\pi\times$5 GHz, $\delta_m = 2\pi\times$2 GHz. (d-f) High-efficiency state in the over-pumped regime. $2\alpha_A=\alpha_C=2\pi\times$8 GHz. $\delta_m=2\pi\times$2 GHz. (g-i) High-efficiency state at the presence of photorefractive effect. $\alpha_A=\alpha_C=2\pi\times$5 GHz, $\delta_m = 2\pi\times$10 GHz.}
	\label{figTime}
\end{figure}

In this section, we present the temporal evolution of the SHG process under the optimal conditions, which further demonstrates the stability of the proposed optimal conditions. The evolution of the SFG process is not shown since the stability conditions are trivial in the weak signal regime. The evolutions without the parasitic effects are calculated by solving the degenerate version of Eqs. (1 - 3) using a fourth order Runge-Kutta method. The time evolution of SHG with the photorefractive effect is modeled by,
\begin{align}
	&\frac{dA}{dt} = -\frac{\alpha_A}{2}A - i(\Delta_{A0} + M_{11}R + M_{13}S)A + i\omega_A \kappa A^\ast C  + \sqrt{\frac{\theta_A}{t_R}}A_{\mathrm{in}},\\
	&\frac{dC}{dt} = -\frac{\alpha_C}{2}C - i(\delta_{m0}+2\Delta_{A0}+M_{31}R+M_{33}S+2M_{11}R+2M_{13}S) C + i\omega_A \kappa A^2,\\
	&\frac{dR}{dt} = -\frac{1}{T_p}R + \frac{1}{T_p}|A|^2,\\
	&\frac{dS}{dt} = -\frac{1}{T_p}S + \frac{1}{T_p}|C|^2,
\end{align}
where $R$ and $S$ are the averaged powers corresponding to the slow response of the photorefractive effect, $T_p$ is the charge relaxation time. We use $T_p=$20 ns in our simulation which is shorter than the typical relaxation time of photorefractive effects. However, the stability properties of the system do not change for $T_p$ values much larger than the cavity lifetime. We chose the current value to avoid running the simulation for an excessively long time.

Figures \ref{figTime}(a-c) correspond to the highest-efficiency state of trace (ii) in Fig. 2(a). Figures \ref{figTime}(d-f) correspond to trace (iii) in Fig. 2(a) with a pump power at 100 mW. A gradual tuning as shown in Fig. \ref{figTime}(d) is required to reach the high-efficiency state. If the detuning is set to the optimal value with an empty cavity, the systems settles to a low efficiency state. When the photorefractive effect is considered, the highest efficiency can be reached at two combinations of pump powers and detunings as Eq. (15) is quadratic. Figures \ref{figTime}(g-i) correspond to the highest-efficiency state with the lower pump power in trace (iii) of Fig. 3(a). As a large detuning is required to compensate for the photorefractive effect, a gradual tuning is required to stay in the high-efficiency branch. A larger relaxation time requires a slower tuning rate to achieve the same performance.

\section{Frequency stability of the high-efficiency states for SHG}
In this section, we show that cavity-based SHG process with high pump powers can lead to optical parametric oscillation (OPO), however, the optimal condition proposed in this paper, with suitable dispersion, can support high pump power and conversion efficiency without exciting OPO.

A general analysis of OPO operation from cascaded SHG is shown in \cite{Leo_PRA_2016}. The system can be modeled by including dispersion term in Eqs. (1-3). For cascaded SHG, the equations are simplified into,
\begin{align}
	&\frac{dA}{dt} = -\frac{\alpha_A}{2}A - i\Delta_A A -i\frac{\beta_2^{(A)}L}{2t_R}\frac{\partial^2}{\partial s^2}A + i\omega_A \kappa A^\ast C  + \sqrt{\frac{\theta_A}{t_R}}A_{\mathrm{in}},\label{S_SFG_OPO_1}\\
	&\frac{dC}{dt} = -\frac{\alpha_C}{2}C - i(\delta_m+2\Delta_A) C -\frac{\Delta\beta_1L}{t_R}\frac{\partial}{\partial s}C -i\frac{\beta_2^{(C)}L}{2t_R}\frac{\partial^2}{\partial s^2}C + i\omega_A \kappa A^2,\label{S_SFG_OPO_2}
\end{align}
where $s$ is the fast time, $\Delta\beta_1$ corresponds to the group velocity difference, and $\beta_2^{(A)}$ and $\beta_2^{(C)}$ are the GVD parameters of the pump and SH fields, respectively. The normalized forms can be written as,
\begin{align}
	&\frac{da}{d\tau} = -a - i\delta a -i\eta_1\frac{\partial^2}{\partial\phi^2}a+ ia^\ast c +F,\label{S_SHG_OPO_Norm_1}\\
	&\frac{dc}{d\tau} = -\alpha c - i\sigma c - i2\delta c -d\frac{\partial}{\partial\phi}c-i\eta_2\frac{\partial^2}{\partial\phi^2}c + ia^2,\label{S_SHG_OPO_Norm_2}
\end{align}
where $\eta_1 = \mathrm{sign}[\beta_2^{(A)}]$, $\eta_2 = \frac{\beta_2^{(C)}}{|\beta_2^{(A)}|}$, $\phi=\sqrt{\frac{\alpha_At_R}{|\Delta\beta_2^{(A)}|L}}$, and $d = 2\Delta\beta_1\sqrt{\frac{L}{\alpha_At_R|\beta_2^{(A)}|}}$. As an example, bulk lithium niobate crystal with a 1550 nm input pump polarized along the z axis has $\beta_2^{(A)} = 99$ ps$^2$/km, $\beta_2^{(C)} = 389$ ps$^2$/km, and $\Delta\beta_1 = 3.02 \times 10^{-10}$ s/m, which corresponds to $\eta_1 = 1$, $\eta_2 = 4$, and $d = 89$. Whether a homogeneous steady state provides net parametric gain to its sideband modes can be studied by providing a smaller perturbation at the sideband frequency. As shown in \cite{Leo_PRA_2016}, the net gain of a sideband frequency $\Omega$ corresponds to the real part of $\lambda$, where $\lambda$ satisfies,
\begin{align}
	\lambda^4 + c_3(\Omega)\lambda^3+c_2(\Omega)\lambda^2+c_1(\Omega)\lambda+c_0(\Omega) = 0. \label{S_OPO_Gain}
\end{align}
The coefficients are defined as,
\begin{align}
	&c_0(\Omega) = 4|a|^2\left(|a|^2+\bar{\alpha}-\bar{\delta}\bar{\sigma}\right)+\left(1+\bar{\delta}^2-|c|^2\right)\left(\bar{\alpha}^2+\bar{\sigma}^2\right),\\
	&c_1(\Omega) = 4|a|^2\left(1+\bar{\alpha}\right)+2\bar{\alpha}\left(1+\bar{\delta}^2+\bar{\alpha}\right)-2|c|^2\bar{\alpha}+2\bar{\sigma}^2,\\
	&c_2(\Omega) = 4|a|^2-|c|^2+1+\bar{\delta}^2+\bar{\sigma}^2+\bar{\alpha}\left(4+\bar{\alpha}\right),\\
	&c_3(\Omega) = 2\left(1+\bar{\alpha}\right),
\end{align}
where $\bar{\alpha} = \alpha+id\Omega$, $\bar{\delta} = \delta-\eta_1\Omega^2$, $\bar{\sigma} = \sigma+2\delta-\eta_2\Omega^2$. When $\Omega = 0$, these equations can be reduced to the temporal stability equation which can be solve analytically, as we have shown in the previous two sections. For general $\Omega$, Eqn. (\ref{S_OPO_Gain}) can be solved numerically and we explore a large range of parameters in this section. There exist several symmetries that can reduce the size of the parameter space. First, the gain remains the same if we flip the sign of $d$ or $\Omega$. This is because the solutions of Eq. (\ref{S_OPO_Gain}) changes from $\lambda$ to $\lambda^\ast$ if we change $d$ to $-d$ or $\Omega$ to $-\Omega$, which does not change the gain. Second, if we flip the signs of $\delta$, $\sigma$, $\eta_1$, and $\eta_2$ simultaneously, the gain remains the same. This is because such changes do not alter the steady state powers or the coefficients of Eq. (\ref{S_OPO_Gain}).

\begin{figure}
	\centering
	\includegraphics[width=\textwidth]{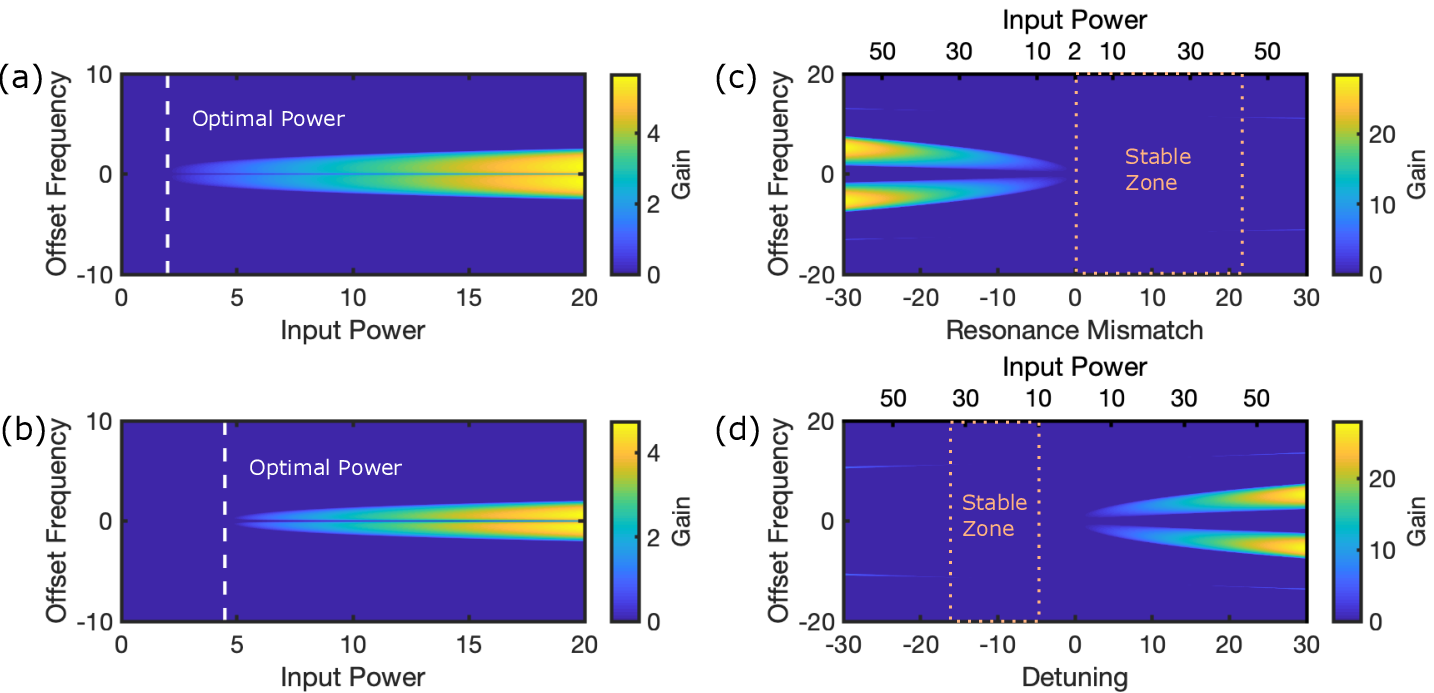}
	\caption{Simulated parametric gain for different regimes of SHG operation. For all simulations $\alpha=1$, $\eta_1=1$, $\eta_2=4$, and $d=60$. (a) Perfectly matched configuration with $\sigma=0$ and $\delta=0$. Parametric gain appears at a power level slightly above the optimal power. (b) Resonance mismatched configuration $\sigma = 2$ and $\delta = -2$. $\delta$ is chosen such that maximum efficiency is reached at the optimal power. (c) Optimal condition in the saturation regime. $\sigma$, $\delta$ and $F$ are jointly varied to maintain the high-efficiency state. (d) Optimal condition in the overpumped regime with $\sigma = 0$. $\delta$ and $F$ are jointly varied to maintain the high-efficiency state. The stable zone is drawn for $F>10$ which is the starting point of the overpumped regime.}
	\label{figGain}
\end{figure}

First we study the perfectly matched configuration $\sigma=0$ and $\delta=0$, where we explore the relation between gain spectrum and input power. We find that with large walkoff ($d>$20), the gain spectrum show very weak dependence on the values of $d$ and $\eta_2$ or the sign of $\eta_1$ (which can only be $\pm$ 1). Moreover, the value of $\alpha$ shifts the entire plot horizontally in which the net gain always starts to appear at an input power slightly above the optimal power given in Eq. (\ref{S_SHG_Opt_2}). We have simulated all combinations of $\alpha \in \{0.5, 1, 2, 3, 4\}$, $\eta_1 \in \{\pm1\}$, $\eta_2 \in \{\pm8, \pm6, \pm4, \pm1, \pm0.5\}$, and $d \in \{20, 40, 60, 80, 100\}$, which confirms the observation. As an example, Fig. \ref{figGain}(a) shows the gain spectrum with $\alpha=1$, $\eta_1 = 1$, $\eta_2=4$, and $d=60$, in which parametric oscillation can occur for an input power $F>$ 2. Thus the perfectly matched $\chi^{(2)}$ microresonator can function as an efficient SHG device up to its optimal input power but cannot function for higher power due to the onset of parametric oscillation. 

When detuning is applied, the gain spectrum resembles that of a $\chi^{(3)}$ OPO. In general, for $\delta < 0$ ($\delta > 0$) the system exhibits a positive (negative) $\chi^{(3)}$ behavior, which shows parametric gain with anomalous (normal) GVD at the pump frequency. However, we notice that additional gain can occur depending on the walkoff and the GVD at the SH frequency. Figure \ref{figGain}(c) shows the gain spectrum as a function of resonance mismatch $\sigma$ under the optimal conditions in the saturation regime, with $\alpha=1$, $\eta_1=1$, $\eta_2=4$ and $d=60$. We vary the resonance mismatch $\sigma$ after which $F$ and $\delta$ are determined by Eqs. (\ref{S_SHG_Opt_1}-\ref{S_SHG_Opt_2}). The two thick lobes correspond to the parametric gain of an effective $\chi^{(3)}$ process, which show weak dependence on the walkoff $d$ and the GVD of the SH frequency $\eta_2$ for $d>30$. However, 4 additional narrow lobes of gain are also present which can limit the applicability of the optimal SHG condition. These extra lobes are pushed to higher resonance mismatch/input power regimes with higher walkoff or lower $\eta_2$. With $d=89$ and $\eta_2=4$, the narrow lobes are absent in the simulation window, thus, in this case, the optimal SHG state is stable as long as the GVD and the detuning have opposite signs at the pump frequency. This behavior is confirmed by simulations covering all combinations of $\alpha \in \{0.5, 1, 2, 3, 4\}$, $\eta_1 \in \{\pm1\}$, $\eta_2 \in \{\pm8, \pm6, \pm4, \pm1, \pm0.5\}$, and $d \in \{20, 40, 60, 80, 100\}$. The optimal condition with resonance mismatch also shows more tolerance of pump power variations. As shown in Fig. \ref{figGain}(b), the difference between the optimal power level and the onset of parametric gain is larger than that in Fig. \ref{figGain}(a). For very small walkoff, the gain spectrum is more complicated and show strong dependence on the values of $\eta_2$, $\alpha$, and $d$. For most SHG devices, $d$ is large due to the large separations of the pump and SH fields. For cases where small walkoff is unavoidable, one can use Eq. (\ref{S_OPO_Gain}) to determine the stable zone.

Similar behavior is also observed in the overpumped regime. We show an example of $\alpha=1$, $\eta_1=1$, $\eta_2=4$,  $d=60$, and $\sigma=0$. We vary $\delta$, after which $F$ is given in Eq. (\ref{S_SHG_Over_Opt}). The two main lobes correspond to an effective $\chi^{(3)}$ process with weak dependence on $d$ and $\eta_2$ for $d>30$. The four narrow lobes depend strongly on $d$ and $\eta_2$, and can be pushed toward higher power regimes for large $d$ or small $\eta_2$. We note that for $\sigma = 0$ the overpumped regime corresponds to $F>10$ which is the starting point of our labeling of the stable zone [Fig. \ref{figGain}(d)].

\end{document}